\documentclass[prb,aps,amsmath,amssymb,showpacs,twocolumn]{revtex4-1}
\usepackage{amssymb,ulem}
\usepackage{amsthm}
\usepackage{amsmath}
\usepackage{natbib}
\usepackage[
            pdfstartview=FitH,
            CJKbookmarks=true,
            bookmarksnumbered=true,
            bookmarksopen=true,
            colorlinks,
            pdfborder=001,
            linkcolor=blue,
            anchorcolor=blue,
            citecolor=blue
            ]{hyperref}
\usepackage{amssymb,graphicx,float,dcolumn,bm,subfigure,color}
\usepackage[titletoc,title]{appendix}
\usepackage[dotinlabels]{titletoc}
\usepackage{braket}
\usepackage{nicefrac}

\newcommand{\be}{\begin{equation}}
\newcommand{\ee}{\end{equation}}

\newcommand{\ba}{\begin{eqnarray}}
\newcommand{\ea}{\end{eqnarray}}

\renewcommand{\vec}[1]{{\textbf{\textit{#1}}}}

\begin{document}
\title{Kohn-Sham Theory of the Fractional Quantum Hall Effect}
\author{Yayun Hu}
\author{J. K. Jain}
\affiliation{Physics Department, 104 Davey Laboratory, Pennsylvania State University, University Park, PA 16802}

\begin{abstract}
We formulate the Kohn-Sham equations for the fractional quantum Hall effect by mapping the original electron problem into an auxiliary problem of composite fermions that experience a density dependent effective magnetic field. Self-consistent solutions of the KS equations demonstrate that our formulation captures not only configurations with non-uniform densities but also topological properties such as fractional charge and fractional braid statistics for the quasiparticles excitations. This method should enable a realistic modeling of the edge structure, the effect of disorder, spin physics, screening, and of fractional quantum Hall effect in mesoscopic devices.
\end{abstract}

\maketitle

The Kohn-Sham density functional theory (KS-DFT) uses the electron density to construct a single particle formalism that incorporates the complex effects of many-particle interactions through a universal exchange correlation function~\cite{Giuliani08}.  It is an invaluable tool for treating systems of interacting electrons spanning the disciplines of physics, chemistry, materials science and biology. 
Very little work has been done~\cite{Ferconi95,Heinonen95,Zhao17} toward applying this method to the FQHE~\cite{Tsui82}, which is one of the most remarkable manifestations of interelectron interactions~\cite{Laughlin83,Jain07}. The reasons are evident. 
To begin with, even though the KS-DFT is in principle exact, its accuracy, in practice, is dictacted by the availability of exchange correlation (xc) potentials, and it works best when the xc contribution is small compared to the kinetic energy. In the FQHE problem, the kinetic energy is altogether absent (at least in the convenient limit of very high magnetic fields) and the physics is governed entirely by the xc energy. A more fundamental impediment is that, by construction, the KS-DFT eventually obtains a single Slater determinant solution, whereas the ground state for the FQHE problem is an extremely complex, filling factor-dependent wave function that is not adiabatically connected to a single Slater determinant. In particular, a mapping into a problem of non-interacting electrons in a KS potential will produce a ground state that locally has integer fillings, whereas nature displays preference for certain fractional fillings. Finally, a mapping into a system of weakly interacting electrons will also fail to capture topological features of the FQHE, such as fractional charge and fractional braid statistics for the quasiparticles~\cite{Laughlin83,Halperin84,Arovas84}. At a fundamental level, these difficulties can be traced back to the fact that the space of ground states in the lowest Landau level (LLL) is highly degenerate for non-interacting electrons, and the interaction causes a non-perturbative reorganization to produce the FQHE. We note here that the application of KS-DFT to ``strictly correlated electrons" is in general an important problem and has previously been considered in other contexts~\cite{Seidl99,Seidl99b,Gori-Giorgi09,Malet12}.   

To make progress, we exploit the fact that the strongly interacting electrons in the FQHE regime turn into weakly interacting composite fermions, namely bound states of electrons and an even number ($2p$) of quantum vortices~\cite{Jain89,Jain07}.  This suggests using an auxiliary system of non-interacting composite fermions to construct a KS-DFT formulation of the FQHE, which is the approach we follow in this work. A crucial aspect of our KS theory is that it properly incorporates the long range ``gauge interaction" between composite fermions induced by the Berry phases due to the quantum mechanical vortices attached to them, which is responsible for the topological properties of the FQHE~\cite{Jeon04,Jain07,Zhang14}.

We consider the Hamiltonian for fully spin polarized electrons confined to the LLL:
\begin{equation}
\mathcal H=\hat{H}_{\rm ee}+\int d\vec{r} V_{\rm ext}(\vec{r})\hat{\rho}(\vec{r})
\end{equation}
Within the so-called magnetic-field DFT~\cite{Grayce94,kohn04,Tellgren12,Tellgren18b}, the Hohenberg-Kohn (HK) theorem also applies to interacting electrons in the FQHE regime and  implies that
the ground state density and energy can be obtained by minimizing the energy functional
\begin{equation}
E[\rho]=F[\rho]+\int d\vec{r} V_{\rm ext}(\vec{r})\rho(\vec{r}),
\end{equation}
where the HK functional is given by~\cite{Levy79,Lieb83}
\begin{equation}
F[\rho]=\min_{\Psi_{\rm LLL}\rightarrow \rho(\vec{r})} \langle\Psi_{\rm LLL}|\hat{H}_{\rm ee}|\Psi_{\rm LLL}\rangle \equiv E_{\rm xc}[\rho]+E_{\rm H}[\rho].
\end{equation}
(The $B$ dependence of the energy functional has been suppressed for notational convenience). Here $E_{\rm xc}[\rho]$ and $E_{\rm H}[\rho]$ are the xc and Hartree energy functionals of electrons and $\Psi_{\rm LLL}$ represents a LLL wave function.  
The conventional KS mapping into non-interacting electrons is problematic due to the absence of kinetic energy.

We instead map the FQHE into the auxiliary problem of ``non-interacting" composite fermions. Composite fermions' most fundamental property is that they experience an effective magnetic field. In particular, the integer quantum Hall effect of composite fermions at $\nu^*=n$ manifests as the FQHE of electrons at $\nu=n/(2pn\pm 1)$. (The quantities referring to composite fermions are marked by an asterisk below.) Even though we use the term non-interacting, the Berry phases associated with the bound vortices induce a long range gauge interaction between composite fermions, as a result of which they  experience a density dependent magnetic field $B^*(\vec{r})=B- 2\rho(\vec{r})\phi_0$, where $\phi_0=hc/e$ is a flux quantum. We therefore write
\begin{equation}
\left[\frac{1}{2m^*}\left(\vec{p}+\frac{e}{c}\vec{A}^*(\vec{r};[\rho])\right)^2+V_{\rm KS}^*(\vec{r})\right] \psi_{\alpha}(\vec{r})=
\epsilon_{\alpha} \psi_{\alpha}(\vec{r}),
\label{singleCFKSMain}
\end{equation}
where
$V_{\rm KS}^*(\vec{r})$ is the KS potential for composite fermions, $m^*$ is the composite-fermion (CF) mass, and $\nabla \times \textbf{A}^*(\vec{r};[\rho])=B^*(\vec{r})$. 
As a result of the gauge interaction, the solution for any given orbital depends, through the $\rho(\vec{r})$ dependence of the vector potential, on the occupation of all other orbitals. Eq.~\ref{singleCFKSMain} must therefore be solved self-consistently, i.e., the single-CF orbitals $\psi_\alpha(\vec{r})$ must satisfy the condition that the ground state density $\rho(\vec{r})=\sum_{\alpha} c_\alpha |\psi_{\alpha}(\vec{r})|^2$, where $c_\alpha=1$ (0) for the lowest energy occupied (higher energy unoccupied) single-CF orbitals, is equal to the density that appears in the kinetic energy of the Hamiltonian. The energy levels of Eq.~\ref{singleCFKSMain} are Landau-like levels of composite fermions, called $\Lambda$ levels ($\Lambda$Ls).  For the special case of a spatially uniform density and constant $V_{\rm KS}^*$, Eq.~\ref{singleCFKSMain} reduces to the problem of non-interacting particles in a uniform $B^*$. 
Importantly, once a self-consistent solution is found for a given $V^*_{\rm KS}(\vec{r})$, for the corresponding density in the Hamiltonian in Eq.~\ref{singleCFKSMain}, the ground state satisfies, by definition, the self-consistency condition and also the variational theorem, and the standard proof for the HK theorem follows. See Supplementary Materials (SM)~\cite{SI-Hu-2019} for details. We define the CF kinetic energy functional as
\be
T_{\rm s}^*[\rho]=\min_{\Psi\rightarrow \rho}\langle\Psi |  \frac{1}{2m^*}\sum_{j=1}^N\left(\vec{p}_j+\frac{e}{c}\vec{A}^*(\vec{r}_j;[\rho])\right)^2 | \Psi\rangle
\ee
where we perform a constrained search over all single Slater determinant wave functions $\Psi$ that correspond to the density $\rho(\vec{r})$, following the strategy of the generalized KS scheme~\cite{Seidl96,SI-Hu-2019}.

The next key step is to write $F[\rho]=T_{\rm s}^*[\rho]+E_{\rm H}[\rho]+E^*_{\rm xc}[\rho]$.
Such a partitioning of $F[\rho]$ can, in principle, always be made given our assumptions, but is practically useful only if the $T_{\rm s}^*[\rho]$ and $E_{\rm H}[\rho]$ capture the significant part of $F[\rho]$, and the remainder $E^*_{\rm xc}[\rho]$, called the exchange-correlation energy of composite fermions, makes a relatively small contribution. This appears plausible given that the CF kinetic energy term captures the topological aspects of the FQHE, and also because the model of weakly interacting composite fermions has been known to be rather successful in describing a large class of experiments.

Minimization of the energy
$
E[\rho]=T_{\rm s}^*[\rho]+E_{\rm H}[\rho]+E^*_{\rm xc}[\rho]+\int d\vec{r} V_{\rm ext}(\vec{r})\rho(\vec{r})
$
with respect to $\rho(\vec{r})=\sum_{\alpha} c_\alpha |\psi_{\alpha}(\vec{r})|^2$, subject to the constraint $\int d \vec{r} \psi_{\alpha}^*(\vec{r})\psi_{\beta}(\vec{r})=\delta_{\alpha\beta}$, yields~\cite{SI-Hu-2019} Eq.~\ref{singleCFKSMain} with
\begin{equation}
V^*_{\rm KS}[\rho,\{\psi_\alpha\}]=V_{\rm H}(\vec{r})+V_{\rm xc}^{\rm  *}(\vec{r})+V_{\rm ext}(\vec{r})+V^*_{\rm T}(\vec{r}),
\label{VKS*}
\end{equation}
where $V_{\rm H}(\vec{r})= \delta E_{\rm H}/\delta \rho(\vec{r})$ and $V_{\rm xc}^{\rm  *}(\vec{r})=\delta E_{\rm xc}^*/\delta \rho(\vec{r})$ are the Hartree and CF-xc potentials. The non-standard potential
\be
V^*_{\rm T}(\vec{r})=\sum_{\alpha} c_\alpha \langle\psi_{\alpha}|\frac{\delta T^*}{\delta \rho(\vec{r})}|\psi_{\alpha}
\label{VT*}
\rangle
\ee
with $T^*=\frac{1}{2m^*}\left(\vec{p}+\frac{e}{c}\vec{A}^*(\vec{r};[\rho])\right)^2$ arises due to the density-dependence of the CF kinetic energy.  $V^*_{\rm T}$ describes the change in $T_{\rm s}^*$ to a local disturbance in density for a fixed choice of the KS orbitals.  Eqs.~\ref{singleCFKSMain}-\ref{VT*} define our KS equations. Because $V^*_{\rm T}(\vec{r})$ depends not only on the density but also on the occupied orbitals, we are actually working with what is known as the ``orbital dependent DFT"~\cite{Kummel08}.

Having formulated the CF-DFT equations, we now proceed to obtain solutions for some representative cases. 
The primary advantage of our approach is evident without any calculations. Take the example of a uniform density FQHE state at $\nu=n/(2pn\pm 1)$. It is an enormously complicated state in terms of electrons, but maps into the CF state at filling factor $\nu^*=n$ with a spatially uniform magnetic field, thereby producing the correct density without any fine tuning of parameters or averaging.  For non-uniform densities, the state of non-interacting composite fermions will produce configurations where composite fermions {\it locally} have $\nu^*\approx n$, which corresponds to an electronic state where the local filling factor is $\nu\approx n/(2pn\pm 1)$, which is a reasonable description, and certainly a far superior representation of the reality than any state of non-interacting electrons.

For a more quantitative treatment we need a model for the xc energy. To this end, we begin by making the local density approximation (LDA) to write $E_{\rm xc}^*[\rho]=\int d\vec{r} \epsilon^*_{\rm xc}[\rho(\vec{r})]\rho(\vec{r})$, where $\epsilon^*_{\rm xc}[\rho]$ is the xc energy per CF. In the following, we express all lengths in units of the magnetic length $l_B=\sqrt{\hbar c/eB}$ and energies in units of $e^2/\epsilon l_B$, where $\epsilon$ is the dielectric constant of the background. The density is related to the local filling factor as $\nu(\vec{r})=\rho(\vec{r}) 2\pi l_B^2$.  We take the model $\epsilon^*_{\rm xc}[\rho]=a\nu^{1/2}+(b-f/2) \nu+ g$,  with $a=-0.78213$, $b=0.2774$, $f=0.33$, $g=-0.04981$. The first term $a\nu^{1/2}$ is chosen to match with the known classical value in the limit $\nu\rightarrow 0$~\cite{Bonsall77}, and the remaining terms ensure that the sum of $\epsilon^*_{\rm xc}$ and CF kinetic energy accurately reproduces the known electronic xc energies at $\nu=n/(2n+1)$~\cite{SI-Hu-2019}. Although optimized for $\nu=n/(2n+1)$ and small $\nu$, we will uncritically assume this form of $\epsilon^*_{\rm xc}(\nu)$ for all $\nu$.  Our aim in this work is to establish the proof-of-principle validity and the applicability of our approach and its ability to capture topological features; a more extensive search for the most optimal $E_{\rm xc}^*$ is left for future work. The topological properties we focus on in this paper are largely robust against the precise form of the xc energy. The xc potential is given by $V^*_{\rm xc}=\delta E_{\rm xc}^*/\delta \rho(\vec{r})=\frac{3}{2}a\nu^{1/2}+(2b-f)\nu-g$.

\begin{figure}[t]
\includegraphics[width=\columnwidth]{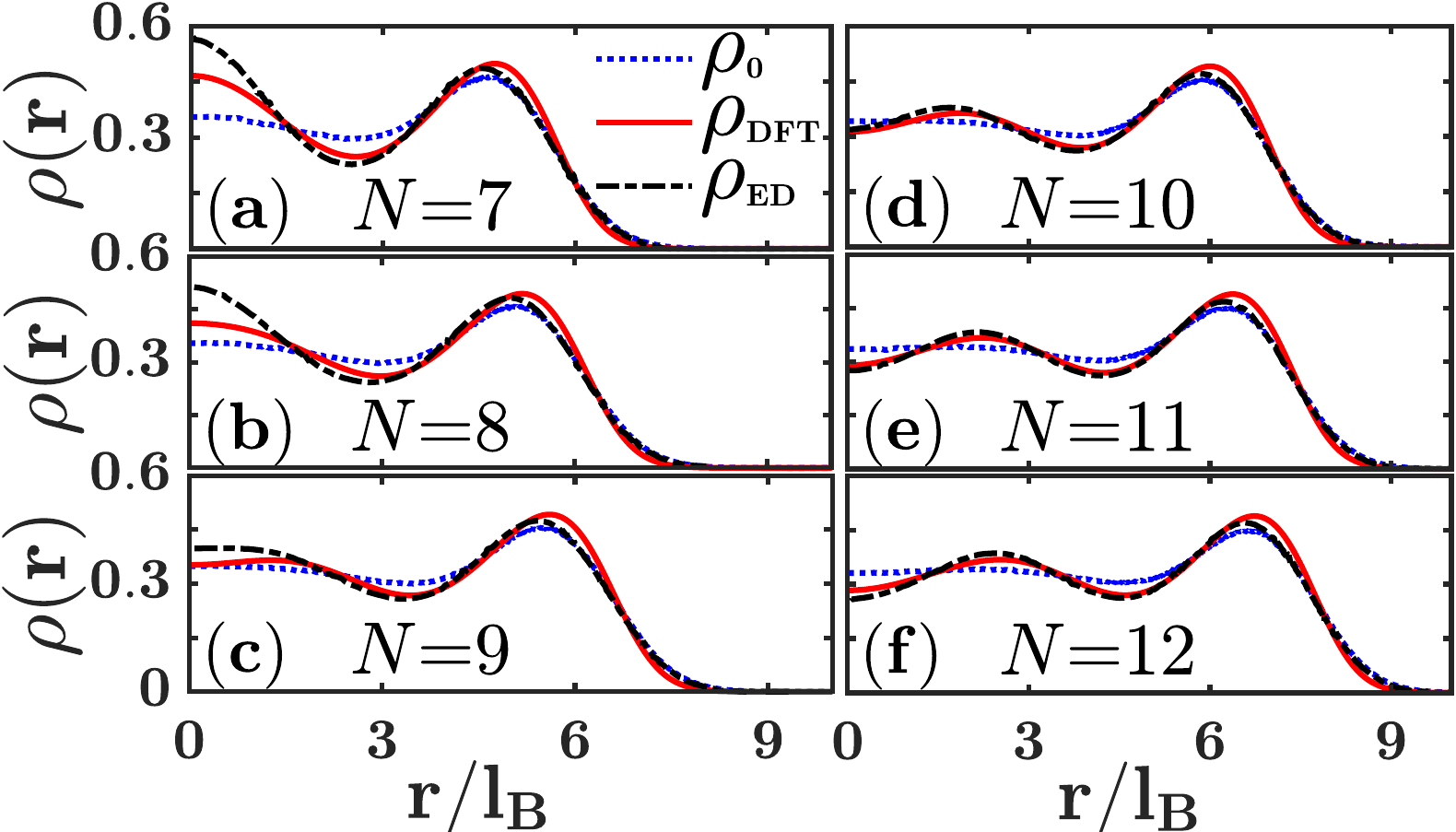}
\caption{Density profile for 1/3 droplets. This figure shows the density of a system of $N$ composite fermions. $\rho_0$ is the density for Laughlin's 1/3 wave function~\cite{Laughlin83}, and $\rho_{\rm ED}$ is obtained from exact diagonalization (ED) of the Coulomb interaction at total angular momentum $L_{\rm total}=3N(N-1)/2$~\cite{Tsiper01}. The density $\rho_{\rm DFT}$ is calculated from the solution of the KS equations for composite fermions in an external potential produced by a uniform positively charged disk of radius $R$ so that $\pi R^2\rho_b=N$. The total angular momentum of the CF state is $L^*_{\rm tot}$, which is related to the total angular momentum of the electron state by $L_{\rm tot}=L^*_{\rm tot}+N(N-1)$~\cite{Jain95}. The CF-DFT solution produces $L^*_{\rm tot}=N(N-1)/2$, which is consistent with $L_{\rm tot}=3N(N-1)/2$. 
All densities are quoted in units of $(2\pi l_B^2)^{-1}$, the density at $\nu=1$. We take $\rho_b=1/3$.}\label{MCDFTcompareMain}
\end{figure}

In our applications below, we will consider $N$ electrons in a potential $V_{\rm ext}(\vec{r})=-\int d^2 \vec{r}' {\rho_{\rm b}(\vec{r}') \over \sqrt{|\vec{r}-\vec{r}'|^2+d^2}}$ generated by a two-dimensional uniform background charge density $\rho_b=\nu_0/2\pi l_B^2$ distributed on a disk of radius $R_{\rm b}$ satisfying $\pi R_{\rm b}^2 \rho_b=N$ at a separation of $d$ from the plane of the electron liquid. This produces an electron system at filling factor $\nu=\nu_0$ in the interior of the disk. We use $\nu_0=1/3$ and $d/l_B\rightarrow 0$ in our calculations below.  For the vector potential, we assume circular symmetry and choose the gauge $\textbf{A}^*(\vec{r})=\frac{r\mathcal{B}(r)}{2}\mathbf{e}_{\phi}$, with $\mathcal{B}(r)=\frac{1}{\pi r^2}\int_0^r 2\pi r'B^*(r') dr'$.

We obtain self-consistent solutions of Eqs.~\ref{singleCFKSMain}-\ref{VT*} by an iterative process. 
Even though we are interested in the zero temperature limit in this article, we sometimes find it useful to begin with a finite temperature $k_B T\sim 0.1$, and anneal the system to approach successively lower temperatures~\cite{Jones14,SI-Hu-2019}. 

\begin{figure}[t]
\centering\includegraphics[width=\columnwidth]{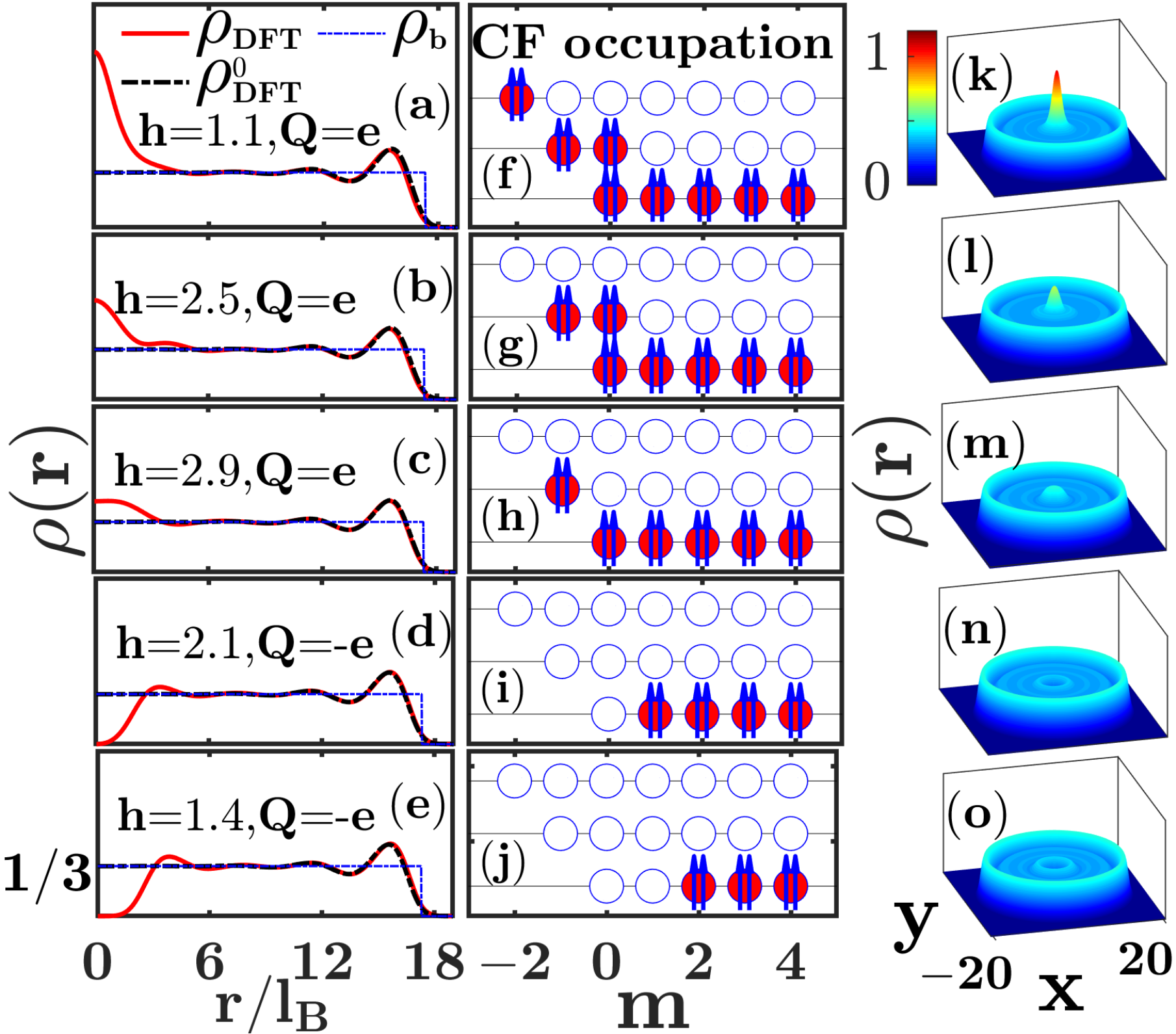}
\centering\includegraphics[width=\columnwidth]{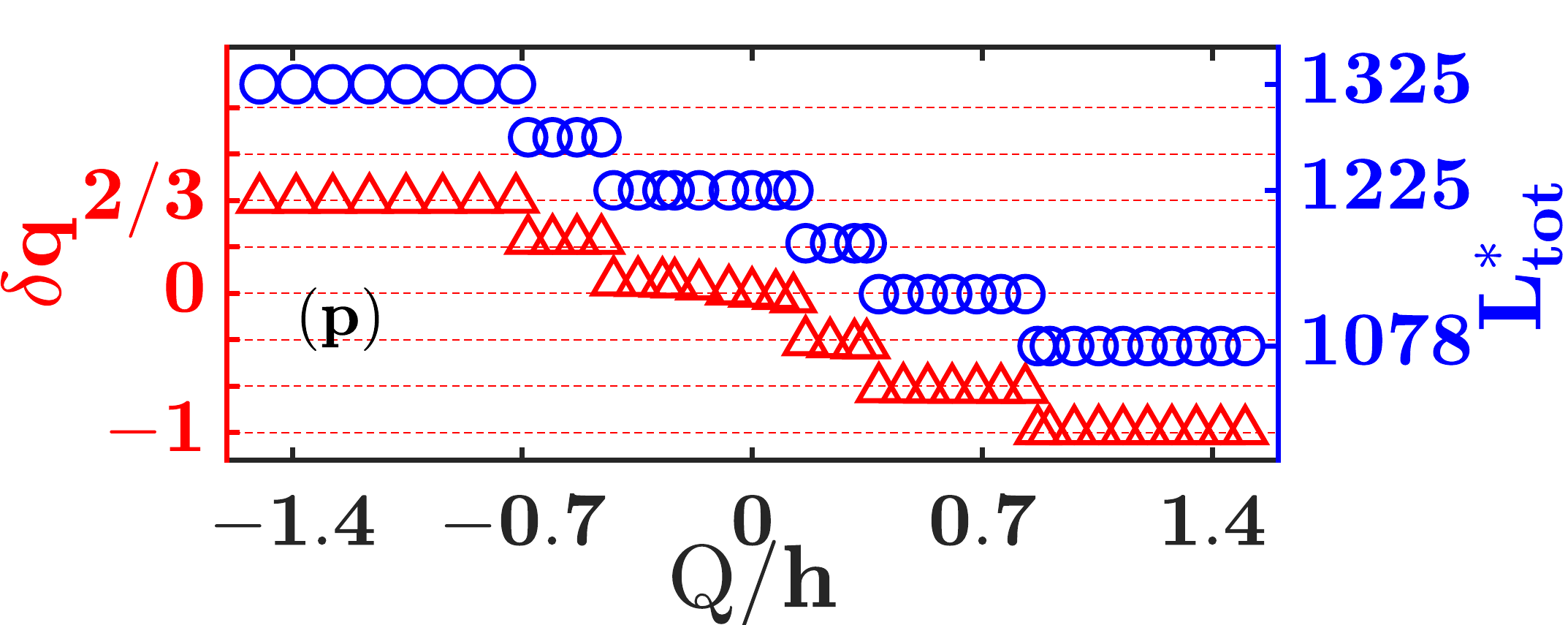}
\caption{Screening and fractional charge. This figure shows how the 1/3 state screens a charged impurity of strength $Q=\pm e$ located at a perpendicular distance $h$ from the origin. The panels {(a)-(e)} and {(k)-(o)} show the self-consistent density $\rho_{\rm DFT}(\vec{r})$. Also shown are $\rho^0_{\rm DFT}(\vec{r})$, the ``unperturbed" density (for $Q=0$), and $\rho_b$, which is the density of the positively charged background. Panels {(f-j)} show the occupation of renormalized $\Lambda {\rm L}$s in the vicinity of the origin; each composite fermion is depicted as an electron with two arrows, which represent quantized vortices. (The single particle angular momentum is given by $m=-n, -n+1, \cdots$ in the $n^{\rm th}$ $\Lambda$L.) The panel {\bf(p)} shows the evolution of the excess charge $\delta q$ and the total CF angular momentum $L^*_{\rm tot}$ as a function of the impurity potential strength at the origin $V_{\textrm{imp}}(r =0)=Q/h$. Change in the charge at the origin is associated with a change in $L^*_{\rm tot}$. The system contains a total of $N=50$ composite fermions. For $h=\infty$, we have $L^*_{\rm tot}=1225$ and $\delta q=0$. For one and two quasiholes, we have $L^*_{\rm tot}=1225$ and 1275, whereas for one, two and three quasiparticles we have $L^*_{\rm tot}=1175$, 1127 and 1078, precisely as expected from the configurations in panels {(f)-(j)}~\cite{Jain95}.}\label{kBT1N50IM}
\end{figure}

As a first application, we consider the density profile of the $\nu_0=1/3$ droplet. Fig.~\ref{MCDFTcompareMain} shows the density profiles calculated from Laughlin's trial wave function~\cite{Laughlin83} as well as that obtained from exact diagonalization at total angular momentum $L=3N(N-1)/2$~\cite{Tsiper01}.  Also shown are the density profiles obtained from the above KS equations. The  density profile from our CF-DFT captures that obtained in exact diagonalization well, especially for $N\geq 10$. Remarkably, it reproduces the characteristic shape near the edge where the density exhibits oscillations and overshoots the bulk value before descending to zero. This qualitative behavior is fairly insensitive to the choice of $V^*_{\rm xc}$, and is largely a result of the self-consistency requirement in Eq.~\ref{singleCFKSMain}~\cite{SI-Hu-2019}. The SM considers other configurations, and also shows that a mean-field approximation is highly unsatisfactory for the density profile.

We next consider screening of an impurity with charge $Q=\pm e$ at a height $h$ directly above the center of the FQHE droplet. The strength of its potential
\begin{equation}\label{Vr}
V_{\rm imp}(\vec{r})=\frac{Q}{\sqrt{|\vec{r}|^2+h^2}}
\end{equation}
can be tuned by varying $h$.
Panels (a)-(e) in Fig.~\ref{kBT1N50IM} show the density $\rho$ for certain representative values of $h$.
It is important to note that the CF orbitals in the self-consistent solution form strongly renormalized $\Lambda {\rm L}s$ (i.e. include the effect of mixing between the unperturbed $\Lambda$Ls). Panels (f)-(j) show the occupation of the $\Lambda {\rm L}s$. The presence of the impurity either empties some CF orbitals from the lowest $\Lambda$L or fills those in higher $\Lambda$Ls. Each empty orbital in the lowest $\Lambda$L corresponds to a charge 1/3 quasihole, whereas each filled orbital in an excited $\Lambda$L to a charge $-1/3$ quasiparticle~\cite{Jain07}. The excess charge is defined as $\delta q=\int_{|r|<r_0} d^2\vec{r} [\rho_0-\rho(\vec{r})]$ in a circular area of radius $r_0=10 l_B$ around the origin.  Panel (p) shows how $\delta q$ and $L^*_{\rm tot}$ change as a function of the potential at the origin $V_{\rm imp}(\vec{r}=0)=-Q/h$.  The excess charge $\delta q$ is seen to be quantized at an integer multiple of $\pm 1/3$.

We finally come to fractional braid statistics. Particles obeying such statistics, called anyons, are characterized by the property that the phase associated with a closed loop of a particle depends on whether the loop encloses other particles. In particular, for abelian anyons, each enclosed particle contributes a phase factor of $e^{i 2\pi \alpha}$, where $\alpha$ is called the statistics parameter. [For non-interacting bosons (fermions), $\alpha$ is an even (odd) integer.] In the FQHE, the quasiparticles are excited composite fermions and quasiholes are ``missing" composite fermions. Let us consider quasiholes of the 1/3 state for illustration. A convenient way to ascertain the statistics parameter within our KS-DFT is to ask how the location of a quasihole in angular momentum $m$ orbital changes when another quasihole is inserted at the origin in the $m=0$ orbital. Let us first recall what is the expected behavior arising from fractional braid statistics. In an effective description, the wave function of a single quasihole in angular momentum $m$ orbital is given by $z^m e^{-|z|^2/4l^{*2}}$ ($z\equiv x-iy$), which is maximally localized at $r_{\rm ex}=(2m)^{1/2}l^{*}=(6m)^{1/2}l_B$, with $l^*=\sqrt{3}l_B$ (as appropriate for $\nu_0=1/3$). When another quasihole is present at the origin, it induces an additional statistical phase factor $e^{i2\pi\alpha}$, where $\alpha$ is the statistics parameter. This changes the wave function of the outer quasihole to $z^{m-\alpha} e^{-|z|^2/4l^{*2}}$, which is now localized at $r_{\rm ex}'=[6(m-\alpha)]^{1/2}l_B$. We now determine $\alpha$ from our KS-DFT formalism.

\begin{figure}[t]
\includegraphics[width=\columnwidth]{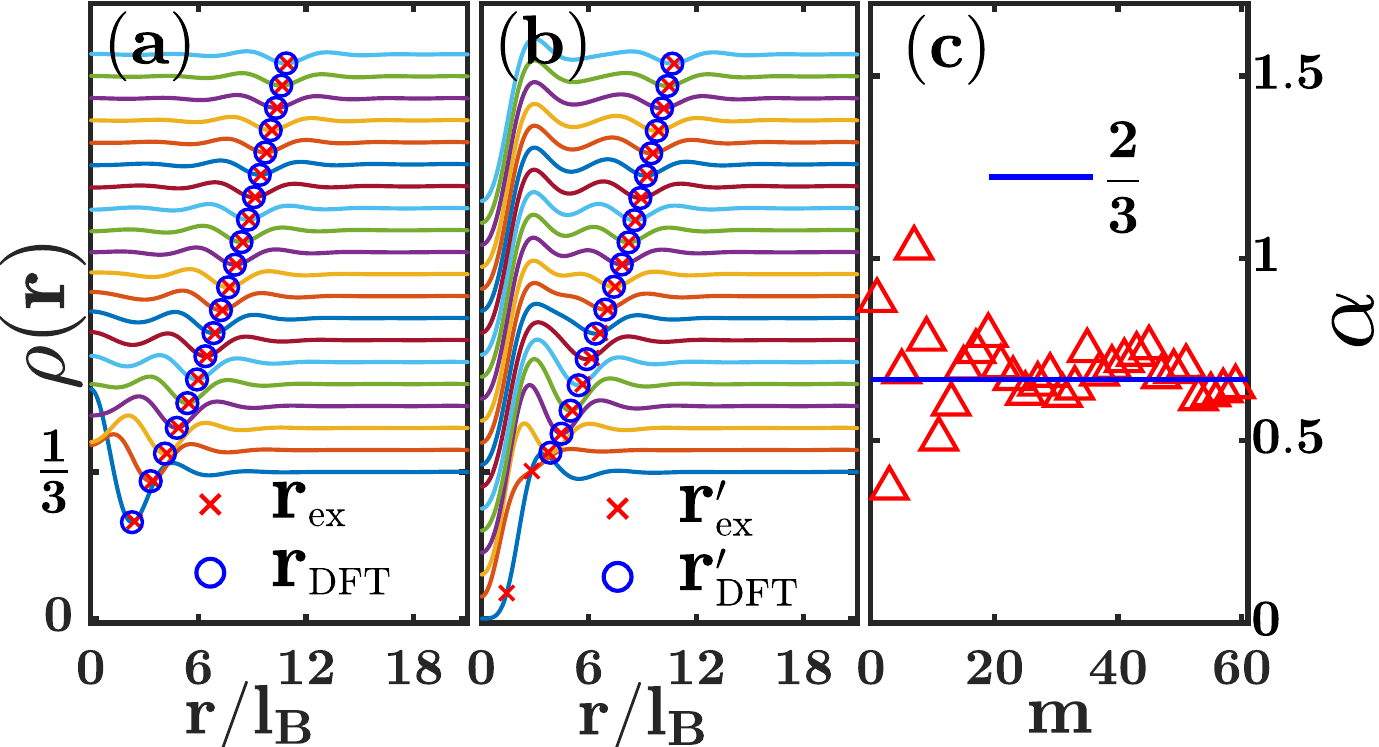}
\caption{Fractional braid statistics. Panel {\bf(a)} shows the electron density for a system with a quasihole in angular momentum $m$ orbital, with $m$ changing from 1 to 20 for the curves from the bottom to the top. (Each successive curve has been shifted up vertically for clarity.) Panel {\bf(b)} shows the same in the presence of another quasihole at the origin. For each $m$, we indicate the expected position of the outer quasihole (red cross) as well as the position obtained from the DFT density determined by locating the local minimum (blue circle). Panel {\bf(c)} shows the calculated statistics parameter $\alpha\equiv (r^2_{\rm{\rm DFT}}-r^{\prime 2}_{\rm{\rm DFT}})/6l_B^2 $. The calculation has been performed for $N=200$ composite fermions at $\nu_0=1/3$.}\label{Braiding}
\end{figure}

A quasihole can be treated in a constrained DFT~\cite{Kaduk12} wherein we leave a certain angular momentum orbital unoccupied. The panels (a) and (b) of Fig.~\ref{Braiding} show the self-consistent KS density profiles of the state with a quasihole in angular momentum $m$, without and with another quasihole in the $m=0$ orbital. The locations of the outer quasihole, $r_{\rm DFT}$ and $r'_{\rm DFT}$, are determined from the minimum in the density. These are in reasonable agreement with the expected positions $r_{\rm ex}$ and $r_{\rm ex}'$ (provided $m> 3$).  More importantly, the calculated statistics parameter $\alpha\equiv (r^2_{\rm{\rm DFT}}-r^{\prime 2}_{\rm{\rm DFT}})/6l_B^2$ is in excellent agreement with the expected fractional value of $\alpha=2/3$~\cite{Halperin84,Jain07} provided that the two quasiparticles are not close to one another, indicating that our method properly captures the physics of fractional braid statistics. The small deviation from 2/3 for large $m$ arises from the fact that the density of the unperturbed system itself has slight oscillations due to the finite system size, which causes a slight shift in the position of the local minimum due to an additional quasihole. Correcting for that effect produces a value much closer to $\alpha=2/3$, as illustrated in the SM~\cite{SI-Hu-2019}.

In conclusion, we have formulated in this article a Kohn-Sham DFT that faithfully captures the topological characteristics of the FQHE state, such as fractional charge, fractional statistics and effective magnetic field. This opens a new strategy for exploring a variety of problems of interest. Aside from the nature of FQHE edges, our approach should allow a quantitative treatment of the effect of smooth disorder, as well as of correction due to Landau level mixing and finite width through appropriate modifications of the xc potential. One can anticipate a generalization of the KS-DFT to paired CF states supporting non-Abelian excitations. Modeling of mesoscopic devices should provide important insight into the optimal conditions for the measurement of fractional statistics through interference experiments (e.g. Ref.~\onlinecite{Nakamura19}).

\textit{Acknowledgement}:
We are grateful to Gerald Knizia, Andre Laestadius, Paul Lammert, Melvyn Levy, and Erik Tellgren for very useful discussions and advice. We acknowledge financial support from the US Department of Energy under Grant No. DE-SC0005042. Y. H. thanks Yang Ge and PSU DFT Cafe for illuminating help, and acknowledges partial financial support from China Scholarship Council.

\pagebreak

{\large{\bf SUPPLEMENTARY MATERIAL}}

\vspace{5mm}

\setcounter{figure}{0}
\setcounter{equation}{0}
\renewcommand\thefigure{S\arabic{figure}}
\renewcommand\thetable{S\arabic{table}}
\renewcommand\theequation{S\arabic{equation}}

The plan of the supplementary section is as follows. In Sec.~\ref{FQHEandCF}, we briefly introduce certain basic facts about composite fermions that are useful for this work. We show how to construct an auxiliary system of composite fermions to derive the KS equations of FQHE in Sec.~\ref{HowToDeriveKSstar}. An alternative way to obtain the KS equations using the generalized Kohn-Sham scheme is discussed in Sec.~\ref{GKSSProof}. We explain the numerical procedure to find the converged solution to the KS equation in Sec.~\ref{HowToCalculate}. In Sec.~\ref{densityComparisonSec}, we show a mean field model of composite fermions is inadequate. Sec.~\ref{RecalibratedStatistics} contains a demonstration of how the accuracy of the statistics parameter $\alpha$ can be improved by eliminating the fluctuations in the background.

\section{Background on FQHE}
\label{FQHEandCF}

We review some basic facts about composite fermions~\cite{Jain89,Jain07} that will be necessary for what follows. For simplicity, we use the disk geometry and assume a rotationally invariant system. The appropriate magnetic vector potential is $\vec{A}=\frac{rB}{2}\vec{e}_{\phi}$ in the symmetric gauge. For a single electron in a uniform magnetic field, the solution to the Schr\"{o}dinger equation:
\begin{equation}\label{ElectronKinetic}
  \hat{T}\phi_{n,m}(\vec{r})=\frac{1}{2m_{e}}\left(\vec{p}+\frac{e}{c}\vec{A}\right)^2\phi_{n,m}(\vec{r})=E_{n,m}\phi_{n,m}(\vec{r})\;,
\end{equation}
is given by
\be\label{LandauWave}
\phi_{n,m}({\vec{r}};l_B)=N_{n,m}\exp\left(-\frac{{r}^2}{4l_B^2}\right)\left({z\over l_B}\right)^{m}L_n^{m}\left({r^2 \over 2l_B^2}\right)\;,
\ee
\be
E_{n,m}=\left( n+{1\over 2}  \right)\hbar\omega_c\;.
\ee
Here the position is denoted by $z=x-iy$ or $\vec{r}=(x,y)$, and $l_B=\sqrt{\hbar c/eB}$ is the magnetic length. The subscript $n=0,1,2,\cdots$ is the Landau level index, $m=-n,-n+1,\cdots$ is the angular momentum index, and $L_n^{m}$ are the associated Laguerre polynomials. $N_{n,m}=\frac{(-1)^n}{\sqrt{2\pi}}\sqrt{\frac{2^nn!}{2^m(m+n)!}}$ is the normalization coefficient. The $z$-component of angular momentum is $\hat{L}_{z}\phi_{n,m}=-m\hbar\phi_{n,m}$.

For the many electron problem, an important parameter is the filling factor $\nu(\vec{r})=2\pi l_B^2 \rho(\vec{r})$, which denotes the number of filled Landau levels (which can in general be spatially varying). For integer fillings $\nu=n$, the ground state of non-interacting electrons has a gap to excitations, and the interaction can be neglected. The FQHE occurs for partially filled LLs, most prominently for electrons in the LLL. (Here, neglecting spin or valley degrees of freedom, we have $\nu<1$.)  For a partially filled LL the ground state has a large degeneracy for non-interacting electrons, and the nature of the interacting ground state is governed entirely by the interaction. The FQHE is a non-perturbative consequence of the interaction, which, no matter how small, opens gaps at certain magic filling factors.

The FQHE is understood in terms of emergent particles called composite fermions. A composite fermion is the bound state of an electron and an even number ($2p$) of quantized vortices, often thought of as an electron with $2p$ flux quanta attached to it (here a flux quantum is defined as $\phi_0=hc/e$). We will specialize in this work to $2p=2$. The composite fermions can be treated as weakly interacting to a good approximation. The density of composite fermions is the same as that of electrons, but composite fermions experience a reduced magnetic field, $B^*=B-2\rho\phi_0=(1-2\nu)B$, where $2\rho\phi_0$ is the flux captured by composite fermions. Composite fermions form Landau-like levels called $\Lambda$ levels ($\Lambda {\rm L}$s) in the reduced magnetic field. The filling factor of composite-fermions, $\nu^*$, is related to electron fillings in Landau levels by $\nu=\nu^*/(2\nu^*+1)$. The effective cyclotron energy is $\hbar \omega_c^*=\hbar(eB^*/m^*c)$ must be determined entirely by the Coulomb energy, and previous studies \cite{Jain07} have shown that a good approximation is $\hbar \omega_c^*=[f/(2\nu^*+1)](e^2/\epsilon l_B)$, with $f=0.33$ for a system with zero thickness.

The phenomenology of the FQHE can be explained in terms of non-interacting composite fermions. In particular, the FQHE at $\nu=n/(2pn\pm 1)$, which are the prominently observed fractions, are understood as the $\nu^*=n$ IQHE of composite fermions\cite{Jain89,Jain07}. In addition, the CF theory also provides accurate microscopic wave functions for interacting electrons. According to the CF theory, basis functions for low-energy states of {\it interacting electrons} in the FQHE regime are constructed as
\be
\label{standard}
\Psi_{\{  n_i,m_i\}  }(B)={\cal P}_{\rm LLL} \Phi_{\{  n_i,m_i\}  }(B)\prod_{j<k}(z_j-z_k)^{2p}\;.
\ee
Here $\Phi_{\{  n_i,m_i\}  }(B)$ denotes a Slater determinant basis constructed at the external magnetic field $B$:
\be
\Phi_{\{  n_i,m_i\}  }(B)={\rm Det}[\phi_{n_i,m_i}(\vec{r}_j,l_B)]
\ee
in which certain set of LL orbitals $\{n_i,m_i\}$ are occupied. The Jastrow factor $\prod_{j<k}(z_j-z_k)^{2p}$ attaches $2p$ quantized vortices to each electron to convert it into a composite fermion. ${\cal P}_{\rm LLL}$ represents projection into the LLL. For uniform density states, this equation relates a basis function $\Phi_{\{n_i,m_i\}  }$ at $\nu^*$ to a correlated basis function $\Psi_{\{n_i,m_i\}  }$ of electrons at $\nu=\nu^*/(2p\nu^*+ 1)$; in particular, for $\nu^*=n$ we have a single Slater determinant for the ground state, which produces a unique wave function for the FQHE state at $\nu=n/(2pn+1)$. For $n=1$, the wave function reduces to Laughlin's wave function. Because we are interested in states with non-uniform densities where arbitrary occupation configuration $\{  n_i,m_i\}$ are allowed, we will not use the filling factor as a label. The density 
\be
\label{electrondensity}
\rho(\vec{r})=\int d\vec{r}_2 \cdots d\vec{r}_N |\Psi_{\{  n_i,m_i\}  }(\vec{r}, \cdots , \vec{r}_N)|^2\;,
\ee
can be calculated using Monte Carlo sampling. 

\section{Derivation of the KS equations}\label{HowToDeriveKSstar}

\subsection{HK Theorem for electrons in FQHE regime}\label{HKandFQHE}

There are two routes to generalizing the HK theorem to systems in a magnetic field. One is called the current-density functional theory (CDFT), and the other  the magnetic-field density functional theory (BDFT). In this paper, we adopt the latter approach, in which the ground state energy and the HK functional are functionals of both the density and the external magnetic field.

Consider the Hamiltonian for fully spin polarized electrons confined to the LLL:
\be
\mathcal H=\hat{H}_{\rm ee}+\int d\vec{r} V_{\rm ext}(\vec{r})\hat{\rho}(\vec{r}).
\label{HFQHE}
\ee
Within the BDFT, the HK theorem also applies to interacting electrons in the FQHE regime, i.e., the external potential $V_{\rm ext}$ can be uniquely determined from the knowledge of the ground state density as well as the external magnetic field $\vec{B}(\vec{r})$ (taken to be spatially uniform in FQHE). This implies that an energy functional
\be
E_{\vec{B}}[\rho]=F_{\vec{B}}[\rho]+\int d\vec{r} V_{\rm ext}(\vec{r})\rho(\vec{r}),
\label{DFT12}
\ee
exists where the HK functional $F_{\vec{B}}[\rho]$ is the minimum interaction energy of wave functions $\Psi_{\rm LLL}$ in the LLL for the given density profile
\begin{equation}\label{FBHK}
  F_{\vec{B}}[\rho]=\min_{\Psi_{\rm LLL}\rightarrow \rho(\vec{r})} \langle\Psi_{\rm LLL}|\hat{H}_{\rm ee}|\Psi_{\rm LLL} \rangle.
\end{equation}
Here $F_{\vec{B}}[\rho]$ depends on the external magnetic field $\vec{B}$. The ground state density can be obtained by minimizing $E_{\vec{B}}[\rho]$.

\subsection{Auxiliary KS systems of non-interacting composite fermions}\label{KSstarSection}

While the HK theorem can be generalized to systems of electrons in the LLL, the KS method fails.
A fundamental assumption for the KS theory is that a choice of KS potential exists for which the density $\rho(\vec{r})$ of noninteracting electrons is the same as that of the true ground state. This assumption is not valid for the FQHE, because a system of non-interacting electrons cannot produce a density that matches the density of the FQHE ground state. For example, for any $V_{\rm KS}$ that is slowly varying at the scale of the magnetic length, the ground state produces regions of integer filling factors rather than fractional filling factors.

We now exploit the fact that the FQHE state maps into a system of weakly interacting composite fermions to construct KS equations for composite fermions. We begin by proving the HK theorem for non-interacting composite fermions while emphasizing certain subtle features.

The Hamiltonian for ``non-interacting" composite fermions is given by
\begin{equation}\label{HStar}
\mathcal H^*[\rho]=\sum_j\frac{1}{2m^*}\left(\vec{p}_j+\frac{e}{c}\vec{A}^*(\vec{r}_j)\right)^2+\int d\vec{r} {V}^*_{\rm KS}(\vec{r})  \hat{\rho}(\vec{r})\;,
\end{equation}
where
\begin{equation}\label{A*equation}
  \nabla \times \vec{A}^*(\vec{r})=\vec{B}^*(\vec{r})=\left[B- 2\rho(\vec{r})\phi_0\right]\vec{e}_{\rm z}\;,
\end{equation}
and $\rho(\vec{r})$ is the electron or CF density. For rotationally symmetric systems, which are what we consider in this article, it is convenient to choose the gauge
\be
\vec{A}^*(\vec{r})=\frac{r\mathcal{B}(r)}{2}\vec{e}_{\rm \phi}\;,
\label{Ar*}
\ee
with
\begin{equation}\label{pseudoB}
  \mathcal{B}(r)=\frac{1}{\pi r^2}\int_0^r 2\pi r'B^*(r') dr'\;.
\end{equation}
We define the problem of a single composite fermion in a spatially non-uniform magnetic field:
\be
\left[ T^* +V^*_{\rm KS}(\vec{r})  \right] \psi_{\alpha}(\vec{r}) = \epsilon_{\alpha} \psi_{\alpha}(\vec{r})\;,
\label{singleCFKS}
\ee
where
\begin{equation}\label{SingleParticleTstar}
  T^*=\frac{1}{2m^*}\left(\vec{p}+\frac{e}{c}\vec{A}^*(\vec{r})\right)^2.
\end{equation}
The solutions of the above equation are the single-CF orbitals $\psi_\alpha(\vec{r})$, where $\alpha$  collectively denotes the $\{n,m\}$ quantum numbers of the composite fermion in the KS potential.
The CF density is given by
\be
\rho(\vec{r})=\sum_{\alpha} c_{\alpha}|\psi_{\alpha}(\vec{r})|^2\;,
\label{rho*rho}
\ee
where $c_{\alpha}$ is the occupation number of the KS orbital labeled by $\alpha$. We choose $c_{\alpha}=1$ for $\alpha=1, 2, \ldots N$ lowest energy orbitals (neglecting degeneracy), and $c_{\alpha}=0$ otherwise. The KS wave function $\Psi$ is a single Slater determinant of the occupied KS orbitals.

Eq.~\ref{singleCFKS} must be solved self consistently.  For any given $V^*_{\rm KS}$, we shall assume that such a self-consistent solution exists. It it important to note that {\it whenever a self consistent solution is found with a given $\rho(\vec{r})$, it is, by construction, the ground state of the corresponding Hamiltonian $\mathcal H^*[\rho]$} (with $\rho(\vec{r})$ treated as a parameter). In the absence of a ground state degeneracy, a given $V^*_{\rm KS}$ produces a unique ground state wave function as well as a unique ground state density.

We next demonstrate the generalization of the HK theorem which states that the KS potential $V^{*}_{\rm KS}$ is also uniquely determined by the ground state density $\rho$. As usual, the theorem is proven by contradiction. Let us consider two KS potentials $V^{*1}_{\rm KS}$ and $V^{*2}_{\rm KS}$ each producing a non-degenerate self-consistent ground sate wave functions $\Psi_1$ and $\Psi_2$ corresponding to densities $\rho_1(\vec{r})$ and $\rho_2(\vec{r})$. Let us assume that $\rho_1(\vec{r})=\rho_2(\vec{r})=\rho$ but $\Psi_1(\vec{r})\neq\Psi_2(\vec{r})$. It follows that

\begin{eqnarray}
  E^*_1&=&\langle \Psi_1| \mathcal H^*_1 |\Psi_1\rangle < \langle\Psi_2| \mathcal H^*_1|\Psi_2\rangle \nonumber \\
   &=& E^*_2+\int d\vec{r} \rho(\vec{r})[V^{*1}_{\rm KS}(\vec{r})-V^{*2}_{\rm KS}(\vec{r})].\label{inequality1}
\end{eqnarray}
Similarly for $E^*_2$ we have,
\begin{equation}\label{inequality2}
    E^*_2<E^*_1-\int d\vec{r}\rho[V^{*1}_{\rm KS}(\vec{r})-V^{*2}_{\rm KS}(\vec{r})].
\end{equation}
Adding Eq.~(\ref{inequality1}) and Eq.~(\ref{inequality2}) leads to a contradiction
\begin{equation}\label{contradiction}
    E^*_1+E^*_2<E^*_2+E^*_1,
\end{equation}
implying that two distinct ground state wave functions, and thus two distinct KS potentials, cannot produce the same density. This proves that a unique KS potential is associated with a given ground state density, i.e. the KS potential is a functional of the ground state density.

The HK theorem implies that $T_{\rm s}^*[\rho]$ defined as:
\be\label{KSstartotalEKrequireHKform}
T_{\rm s}^*[\rho]=\sum_{\alpha}\langle\psi_\alpha | \frac{1}{2m^*}\left(\vec{p}+\frac{e}{c}\vec{A}^*(\vec{r};[\rho])\right)^2 | \psi_\alpha\rangle,
\ee
where Det$[\psi_\alpha(\vec{r}_\beta)]$ is the self-consistent ground state solution of Eq.~(\ref{singleCFKS}), is a functional of the ground state density.
Equivalently, we can write
\be\label{KSstartotalEK}
T_{\rm s}^*[\rho]=\min_{\Psi\rightarrow \rho}\langle\Psi | \frac{1}{2m^*}\sum_{j=1}^N\left(\vec{p}_j+\frac{e}{c}\vec{A}^*(\vec{r}_j;[\rho])\right)^2 | \Psi\rangle
\ee
where $\Psi$ runs over all wave functions that produces density $\rho$ in the constrained search.

\subsection{KS equations}

We now make a key assumption:  {\it The density profile of any physically relevant FQHE ground state can be obtained in the auxiliary non-interacting CF problem by an appropriate choice of $V^*_{\rm KS}(\vec{r})$.} This assumption is certainly valid for a uniform density FQHE ground state, which corresponds to a uniform density solution of Eq.~\ref{HStar} (i.e. for $V^*_{\rm KS}(\vec{r})=$constant). With an appropriate choice of $V^*_{\rm KS}(\vec{r})$ we can produce a density profile that locally contains regions where the density corresponds to CF filling $\nu^*=n$, i.e. to electron filling $\nu=n/(2n+1)$. Furthermore, the renormalization of the CF orbitals due to mixing with higher $\Lambda$Ls allows smooth density profiles.

Our next step is to write the HK functional $F_{\vec{B}}[\rho]$ in Eq.~\ref{DFT12} as
\begin{equation}
\label{NewPartitionFrho}
F_{\vec{B}}[\rho]=T_{\rm s}^*[\rho]+E_{\rm H}[\rho]+E^*_{\rm xc}[\rho]\;.
\end{equation}
Such a partitioning of $F_{\vec{B}}[\rho]$ can, in principle, always be made, but is in practice useful only if the $T_{\rm s}^*[\rho]$ and $E_{\rm H}[\rho]$ capture the significant part of $F_{\vec{B}}[\rho]$, and $E^*_{\rm xc}[\rho]$, called the xc energy of composite fermions, makes a relatively small contribution. We now need to minimize the energy
\be\label{Interpretation}
E_{\vec{B}}[\rho]=T_{\rm s}^*[\rho]+E_{\rm H}[\rho]+E^*_{\rm xc}[\rho]+\int d\vec{r} V_{\rm ext}(\vec{r})\rho(\vec{r})
\ee
for a system with a fixed number of particles with the constraint $\int d \vec{r} \psi_{\alpha}^*(\vec{r})\psi_{\alpha}(\vec{r})-\delta_{\alpha\beta}$.
This requires that
\begin{equation}\label{vanishVariation}
  \frac{\delta E_{\vec{B}}'[\rho]}{\delta \psi_{\alpha}^*(\vec{r})}=\frac{\delta E_{\vec{B}}'[\rho]}{\delta \psi_{\alpha}(\vec{r})}=0
\end{equation}
where
\begin{equation}\label{Constrained}
E_{\vec{B}}'[\rho]=E_{\vec{B}}[\rho]-\sum_{\alpha}\epsilon_{\alpha\beta}\left[\int d \vec{r} \psi_{\alpha}^*(\vec{r})\psi_{\alpha}(\vec{r})-\delta_{\alpha\beta}\right]\;.
\end{equation}
Up to a unitary transformation, this leads to the KS equation Eq.~(\ref{singleCFKS}) with
\begin{equation}\label{VKSstarDisplay}
  V^*_{\rm KS}[\rho,\{\psi_\alpha\}]=V_{\rm H}(\vec{r})+V_{\rm xc}^{\rm  *}(\vec{r})+V_{\rm ext}(\vec{r})+V^*_{\rm T}(\vec{r})\;.
\end{equation}
Here $V_{\rm H}(\vec{r})= \frac{\delta E_{\rm H}}{\delta \rho(\vec{r})}$ is the Hartree potential, $V_{\rm xc}^{\rm *}(\vec{r})=\frac{\delta E^{\rm *}_{\rm xc}[\rho]}{\delta \rho(\vec{r})}$ is the exchange-correlation potential. Compared with the usual Kohn-Sham potential, the KS potential $\hat{V}^*_{\rm KS}$ in Eq.~(\ref{VKSstarDisplay}) has an extra term $V^*_{\rm T}(\vec{r})$ that comes from the density-dependence of the kinetic energy operator. In the disk geometry with a rotational symmetry, we have
\begin{widetext}
\begin{equation}\label{DeltaTStarDef}
V^*_{\rm T}(\vec{r})=\sum_{\alpha}c_{\alpha}\langle\psi_{\alpha}|\frac{\delta T^*}{\delta \rho(\vec{r})}|\psi_{\alpha}\rangle=\sum_{\alpha}c_{\alpha}\int d \vec{r}'\psi^*_{\alpha}(\vec{r}')\big[\frac{1}{m^*}\left(\vec{p}'+\frac{e}{c}\vec{A}^*(\vec{r}')\right)\frac{e}{c}\frac{\delta \vec{A}^*(\vec{r}')}{\delta \rho(\vec{r})}\big]\psi_{\alpha}(\vec{r}'),
\end{equation}
\end{widetext}
where $\frac{\delta \vec{A}^*(\vec{r}')}{\delta \rho(\vec{r})}$ can be replaced by $-\frac{\phi_0\theta(r'-r)}{\pi r'}\vec{e}_{\phi}$ in a rotationally invariant system, and we have used the Coulomb gauge condition $\nabla\cdot\vec{A}^*=0$.
The KS equation Eq.~(\ref{singleCFKS}) must be solved self-consistently. The converged ground state density $\rho$ of the KS system  obtained by occupying the $N$ orbits with the lowest energy, will give the density of the ground state.

\subsection{Thermal KS equations}\label{ThermalKS}

We follow Ref.~[\onlinecite{Jones14}] for constructing the thermal KS equation. At finite temperature, Eq.~(\ref{DFT12}) is replaced by the free energy
\begin{equation}\label{miniGrand}
  \Omega[\rho]=F^{\rm \vec{B},\tau}[\rho]+\int d^3 \vec{r}\rho(\vec{r})(V_{\rm ext}(\vec{r})-\mu)\;,
\end{equation}
where $\tau$ is the temperature, $\mu$ is the Fermi energy, and $F^{\rm \vec{B},\tau}[\rho]$ is defined as
\begin{equation}\label{RealFunctional}
  F^{\rm \vec{B},\tau}[\rho]=\min_{\hat{\Gamma}_{\rm LLL}\rightarrow \rho}\textrm{Tr}~\hat{\Gamma}_{\rm LLL}\big(\hat{H}_{\rm ee}-\tau\hat{S}\big)\;,
\end{equation}
where $\hat{\Gamma}_{\rm LLL}$ is the density matrix within the lowest Landau level and $\hat{S}$ is the entropy operator.

Similarly to the thermal KS scheme, we will now imagine an auxiliary KS system of composite fermions at the same temperature, described by the Hamiltonian Eq.~(\ref{singleCFKS}). The free energy of the auxiliary CF system is
\begin{equation}\label{miniGrandKS}
  \Omega^*_s[\rho]=F^{\rm \mathcal{T}^*,\tau}[\rho]+\int d^3 \vec{r}\rho(\vec{r})(V^*_{\rm KS}(\vec{r})-\mu)\;,
\end{equation}
where $F^{\rm \mathcal{T}^*,\tau}[\rho]$ is called the kentropy in the DFT literature, defined as
\begin{equation}\label{kentropy}
F^{\rm \mathcal{T}^*,\tau}[\rho]=\min_{\hat{\Gamma}\rightarrow \rho}\textrm{Tr}~\hat{\Gamma}\big(\mathcal{T}^*-\tau\hat{S}\big)\;.
\end{equation}

Let us assume that the self-consistent thermal state of the KS system is found. Then we have the density matrix of the KS system $\hat{\Gamma}_{\rm min}=\sum_\alpha c_\alpha|\psi_{\alpha}\rangle\langle\psi_{\alpha}|$, where $\psi_{\alpha}$ is the KS orbital, and $c_\alpha=1/[e^{(\epsilon_{\alpha}-\mu)/k_{\textrm{B}}\tau}+1]$ is the the occupation number. We emphasize that $\hat{\Gamma}_{\rm min}$ is the ground state density matrix in the sense that the density $\rho$ that enters inside the kinetic energy is treated as a fixed parameter, as in the zero temperature case. In the thermal KS system, the density is given by
\begin{equation}\label{KSstardensityagain}
  \rho=\textrm{Tr}~ \hat{\Gamma}_{\rm min}\hat{\rho}=\sum_\alpha c_{\alpha}|\psi_{\alpha}|^2\;.
\end{equation}
The kentropy is
\begin{equation}\label{kentropy}
F^{\rm \mathcal{T}^*,\tau}[\rho]=T_{\rm s}^{*\tau}[\rho]-\tau S^\tau_{\rm s}[\rho]\;,
\end{equation}
where $T_{\rm s}^{*\tau}[\rho]$ is the non-interacting CF kinetic energy
\begin{equation}\label{ExcStar}
  T_{\rm s}^{*\tau}[\rho]=\sum_{\alpha} c_{\alpha}\langle\psi_{\alpha}|T^*|\psi_{\alpha}\rangle\;,
\end{equation}
and $S^\tau_{\rm s}[\rho]$ is the non-interacting entropy
\begin{equation}\label{CFentropy}
  S^\tau_{\rm s}[\rho]=k_{\textrm{B}}\sum_\alpha c_{\alpha}\log(c_\alpha)\;.
\end{equation}

Now we are ready to construct our thermal KS scheme. First, in analogy with Eq.~(\ref{Interpretation}), we can further rewrite Eq.~\ref{miniGrand} as
\begin{equation}\label{Omegadecomposition}
F^{\rm \vec{B},\tau}[\rho]=T^{*\tau}_{\rm s}[\rho]+E_{\rm H}[\rho]+E^{*\tau}_{\rm xc}[\rho]-\tau S^{*\tau}[\rho]\;,
\end{equation}
where the thermal exchange-correlation term is defined as $E^{*\tau}_{\rm xc}[\rho]=F^{\rm \vec{B},\tau}[\rho]-F^{\rm \mathcal{T}^*,\tau}[\rho]-E_{\rm H}[\rho]$. Second, we use the approximation $E^{*\tau}_{\rm xc}[\rho]=E^*_{\rm xc}[\rho]$, since we will focus only on the low temperature limit in this paper. This way, we finally obtain the KS equations at non-zero temperatures, shown in Eqs.~(\ref{singleCFKSintroduction})-(\ref{FDdistribution1}).

\section{Alternative Derivation of the KS equation using the Generalized Kohn-Sham Scheme}\label{GKSSProof}

We refer the reader to Ref.~[\onlinecite{Seidl96}] for generalized Kohn-Sham (GKS) Scheme. In order to apply GKS in the presence of the magnetic field, we again adopt the BDFT approach with the observation that the magnetic field dependence does not influence the validity of the GKS scheme. In GKS scheme, the constrained search is restricted to all single Slater determinants rather than all wave-functions. Let us define the functional 
\begin{equation}\label{Sphi}
  S[\Phi]=\langle\Phi|\mathcal{T}^*[\rho_{\Phi}]|\Phi\rangle\;,
\end{equation}
where $\Phi={\rm Det} [\phi_i],i=1,2,...N$, is a Slater determinant of $N$ orthonormal orbitals, and $\rho_{\Phi}(\vec{r})=\sum_i^{N}|\phi_i(\vec{r})|^2$ is the density of $\Phi$. We also define (using the notation in Ref.~[\onlinecite{Seidl96}])
\begin{eqnarray}
  F^S[\rho] &=&  \min_{\Phi\rightarrow \rho(r)}S[\Phi]=\min_{\{\phi_i\}\rightarrow \rho(r)}S[\{\phi_i\}]\;,\label{FSrho1}\\
  R^S[\rho] &=& F_{\vec{B}}[\rho]-F^S[\rho]\;,\label{FSrho2}
\end{eqnarray}
where $R^S$ is the Hartree-exchange-correlation-energy-like functional, and $F^S[\rho]$ is the non-interacting-kinetic-energy-like functional. Notice that $F^S[\rho]$ is equivalent to $T_s^*[\rho]$ in Eqs.~(\ref{KSstartotalEKrequireHKform}) and (\ref{KSstartotalEK}).
Plugging Eq.~(\ref{FSrho1}) and Eq.~(\ref{FSrho2}) into Eq.~(\ref{DFT12}) gives
\be\label{GKSdecomposition}
E_{\vec{B}}[\rho]=F^S[\rho]+R^S[\rho]+\int d\vec{r} V_{\rm ext}(\vec{r})\rho(\vec{r})\;.
\ee
Rewriting the ground state energy $E_{\rm g}$ as a minimum of the constrained search over all N-representable density $\rho$ yields,
\begin{eqnarray}\label{minimizationGKS}
&&E_{\rm g}=\min_{\rho(r)\rightarrow N}\{E_{\vec{B}}[\rho]\}\\
&&=\min_{\rho(r)\rightarrow N}\{F^S[\rho]+R^S[\rho]+\int dr V_{\rm ext}(r)\rho(r)\}\nonumber\\
&&=\min_{\rho(r)\rightarrow N}\{\min_{\Phi\rightarrow \rho(r)}S[\{\phi_i\}]+R^S[\rho]+\int dr V_{\rm ext}(r)\rho(r)\}\nonumber \\
&&=\min_{\Phi\rightarrow N}\{S[\{\Phi\}]+R^S[\rho[\Phi]]+\int dr V_{\rm ext}(r)\rho([\Phi];r)\}\;.\nonumber
\end{eqnarray}
The minimizing Slater determinant $\Phi={\rm Det} [\phi_i]$ can be determined by the usual Lagrange procedure, and the variation with respect $\phi_j^*$ leads to
\begin{equation}\label{Variation}
\hat{O}^S_j[\{\phi_i\}]\phi_j+V_R\phi_j+V_{\rm ext}\phi_j=\epsilon_j\phi_j, j=1,...,N,
\end{equation}
where
\begin{equation}\label{OSJ}
    \hat{O}^S_j\phi_{j}=\frac{\delta S[\{\phi_i\}]}{\delta \phi_{j}^*}=\bigg[T^*[\rho_{\Phi}]+\sum_{i=1}^N\langle\phi_{i}|\frac{\delta T^*}{\delta \rho(\vec{r})}|\phi_{i}\rangle\bigg]\phi_{j},
\end{equation}
and $V_R=\frac{\delta R^S[\rho]}{ \delta \rho}$. The GKS scheme is up till now formally exact, except that $R^S[\rho]$ is generally unknown and has to be approximated. If we take
\begin{equation}\label{RSinterpretation}
  R^S[\rho]=E_{\rm H}[\rho]+E^*_{\rm xc}[\rho],
\end{equation}
then Eqs.~(\ref{GKSdecomposition}) and (\ref{Variation}) reduce to Eqs.~(\ref{Interpretation}) and (\ref{singleCFKS}).

\begin{figure}[t]
\centering\includegraphics[width=\columnwidth]{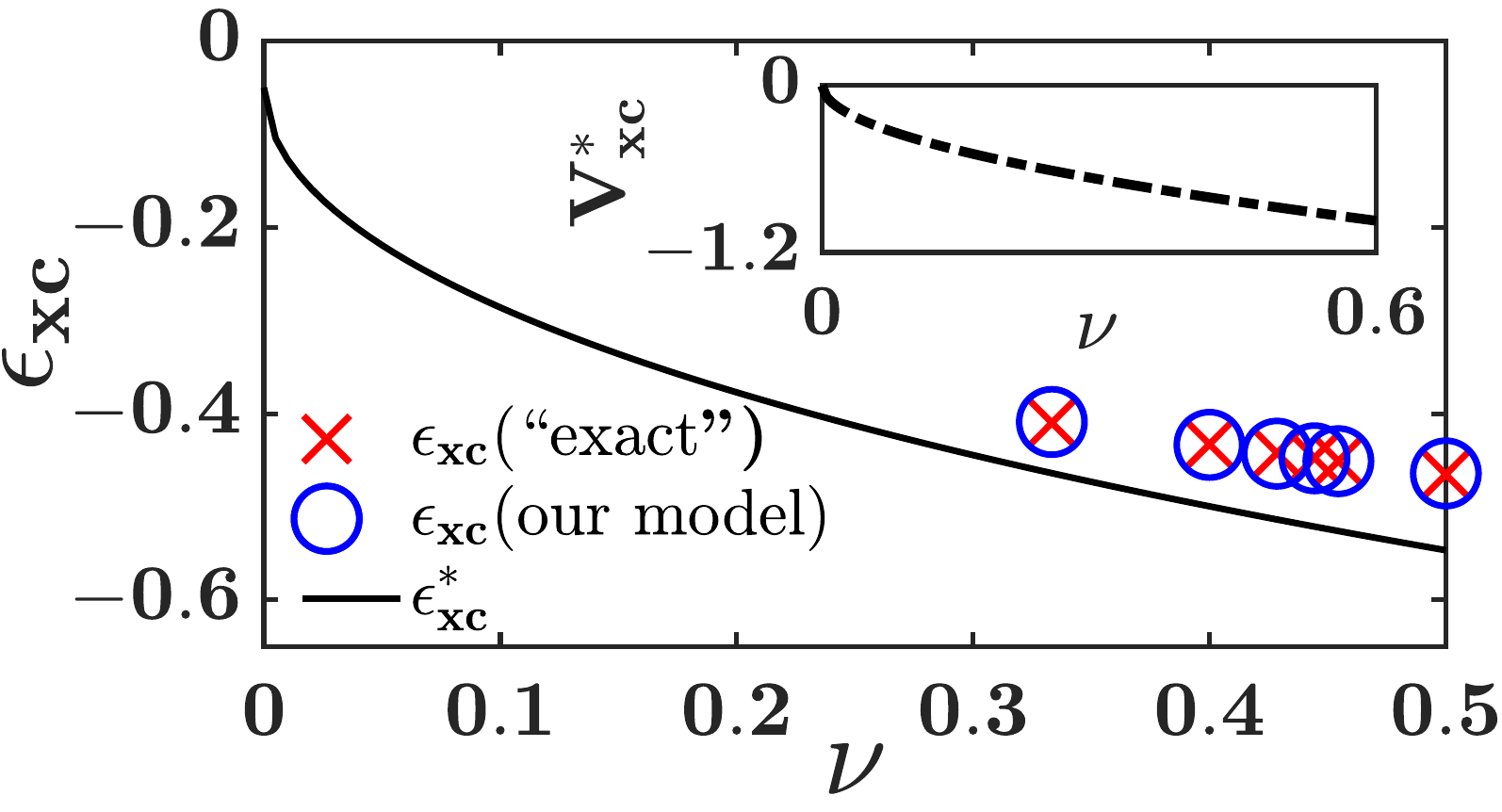}
\caption{{\bf Exchange-correlation energy for composite fermions.} The red crosses mark $\epsilon_{\rm xc}[\nu]$, the xc energy of per electron, at several fractions of the form $\nu=n/(2n+ 1)$, obtained from extrapolation of microscopic CF theory~\cite{Jain07} which is essentially exact. The solid black line shows the CF xc energy $\epsilon^*_{\rm xc}=a\nu^{1/2}+(b-\frac{f}{2})\nu+g$ (parameters $a$, $b$, $f$ and $g$ are given in the text). The blue circles mark $\epsilon_{\rm xc}[\nu]=\epsilon^*_{\rm xc}[\nu]+t^*[\nu]=a\nu^{1/2}+b\nu+g$.  The inset shows the CF xc potential $V_{\rm xc}^*$.}\label{Exc}
\end{figure}

\section{Exchange-correlation potential for composite fermions}

For a quantitative account, we need an accurate model for the xc potential for composite fermions. We focus here on composite fermions with two attached vortices. For the xc energy we make the local density approximation (LDA) to write 
\be
E_{\rm xc}^*[\rho]=\int d\vec{r} \epsilon^*_{\rm xc}[\rho(\vec{r})]\rho(\vec{r}),
\ee 
where $\epsilon^*_{\rm xc}[\rho]$ is the xc energy per CF.   
We determine $\epsilon^*_{\rm xc}$ by requiring that at $\nu=n/(2n+1)$, the relation 
\be 
\epsilon^*_{\rm xc}(\nu)+t^*(\nu)=\epsilon_{\rm xc}(\nu)
\label{exc*}
\ee
is satisfied for uniform densities, where $t^*(\nu)$ is the per-CF kinetic energy and $\epsilon_{\rm xc}(\nu)$ is the exchange correlation energy per electron known from exact diagonalization studies.  The CF kinetic energy is proportional to the cyclotron energy at $\nu=n/(2n\pm 1)$, which is given by 
\be
\hbar \omega^*_c=\hbar {eB^*\over m^*c}= \hbar {eB\over (2n\pm 1)m^*c} \equiv {f \over (2n\pm 1)} {e^2\over \epsilon l_B},
\ee 
where the last step follows because the Coulomb energy is the only energy scale in the problem.  A good approximation for the theoretical transport gaps~\cite{Jain07} with the choice $f=0.33$, which, in turn, corresponds to $m^*=0.079\sqrt{B[T]}m_e$ for parameters appropriate for GaAs, where $m_e$ is the free electron mass and $B[T]$ is the magnetic field in units of Tesla. In our chosen units, the average CF kinetic energy at $\nu^*=n$ is given by
\be 
t^*={n\over 2}\hbar \omega^*_c=f{\nu\over 2}, \;\;\; {\rm for}\;\; \nu={n\over (2n+1)}.
\ee
We take the model 
\be 
\epsilon^*_{\rm xc}[\rho]=a\nu^{1/2}+(b-f/2) \nu+ g,
\ee 
where the first term $a\nu^{1/2}$ is chosen to match with the known classical value in the limit~\cite{Bonsall77} $\nu\rightarrow 0$. As shown in Fig.~\ref{Exc}, Eq.~\ref{exc*} is satisfied with $a=-0.78213, b=0.2774, g=-0.04981$. 
The xc potential is given by $V^*_{\rm xc}=\delta E_{\rm xc}^*/\delta \rho(\vec{r})=\frac{3}{2}a\nu^{1/2}+(2b-f)\nu-g$.

\section{Numerical procedure for KS DFT}\label{HowToCalculate}

It is useful to collect the set of KS equations for composite fermions:
\begin{widetext}
\be
\left[T^*
+V_{\rm H}(\vec{r})+V_{\rm ext}(\vec{r})+V_{\rm xc}^{*}(\vec{r})+V^*_{\rm T}(\vec{r}) \right] \psi_{\alpha}(\vec{r}) = \epsilon_{\alpha} \psi_{\alpha}(\vec{r})\;,
\label{singleCFKSintroduction}
\ee
\end{widetext}
where
\be
T^*=\frac{1}{2m^*}\left(\vec{p}+\frac{e}{c}\vec{A}^*(\vec{r})\right)^2\;,
\ee
\be
\nabla \times \vec{A}^*(\vec{r})=B^*(\vec{r})\vec{e}_{\rm z}=\left[ B- 2\rho(\vec{r})\phi_0\right]\vec{e}_{\rm z}\;,
\ee
\begin{equation}\label{DeltaTStarDefintroduction}
V^*_{\rm T}(\vec{r})=\sum_{\alpha}c_{\alpha}\langle\psi_{\alpha}|\frac{\delta T^*}{\delta \rho(\vec{r})}|\psi_{\alpha}\rangle\;.
\end{equation}
\begin{equation}\label{VH}
  V_{\rm H}(\vec{r})=e^2\int d\vec{r}'\frac{\rho(\vec{r}')}{|\vec{r}-\vec{r}'|}\;.
\end{equation}
\be\label{VXCStarintroduction}
V_{\rm xc}^*(\vec{r})=\frac{3}{2}a\nu^{1/2}(\vec{r})+(2b-f)\nu(\vec{r})-g\;.
\ee

To consider systems at finite temperature $\tau$, we employ the thermal DFT formalism. The set of self-consistent KS equations will have the same form as in Eqs.~(\ref{singleCFKSintroduction})-(\ref{VXCStarintroduction}), with two differences. First, the finite temperature xc energy is different from the one at zero temperature. We will, however, keep using the zero temperature approximation of $V_{\rm xc}^*(\vec{r})$ because we are interested in this work in the $\tau\rightarrow 0$ limit. Second, at finite temperature, the CF density is given by
\be
\rho(\vec{r})=\sum_{\alpha} c_{\alpha}|\psi_{\alpha}(\vec{r})|^2\;,
\label{rho*rhointroduction}
\ee
with 
\be
  c_{\alpha} = \frac{1}{\exp\big([\epsilon_{\alpha}-\mu]/k_\textrm{B}\tau\big)+1}\;, \label{FDdistribution1}
\ee
The chemical potential is determined from $\sum_{\alpha} c_{\alpha}=N$. 
The occupation number $c_{\alpha}$ reduces to either $0$ or $1$ at zero temperature.

\subsubsection{Numerical method to solve the KS equation}

We now outline the numerical procedure to solve the KS Eqs.~(\ref{singleCFKSintroduction}-\ref{FDdistribution1}).We will be assuming rotational symmetry throughout our work, which preserves the angular momentum as a good quantum number. The external potentials considered below, for example the background potential or impurity potential, will be chosen to be rotationally symmetric. Rotational symmetry allows us to write the KS orbital with angular momentum $m_{\alpha}$ as
\begin{equation}\label{variablesepperation}
  \psi_{\alpha}(\vec{r})=\frac{\tilde{R}_{\alpha}(r)}{\sqrt{2\pi r}}\exp(im_{\alpha}\theta).
\end{equation}
We further rewrite $\mathcal H^*$ in the polar coordinate
\begin{widetext}
\be\label{ksStarpolarform}
  \mathcal H^*=\frac{1}{2m^*}\left(\vec{p}+\frac{e}{c}\vec{A}^*(\vec{r})\right)^2+V^*_{\rm KS}=\frac{1}{2m^*}\bigg[-\hbar^2\big(\frac{\partial^2}{\partial r^2}+\frac{1}{r^2}\frac{\partial^2}{\partial \phi^2}+\frac{1}{r}\frac{\partial}{\partial r}\big)
+\frac{\hbar e}{c}\mathcal{B}(r)\frac{\hat{L}_z}{\hbar}+\frac{e^2}{c^2}\frac{r^2\mathcal{B}^2(r)}{4}\bigg]+V^*_{\rm KS}.
\ee
\end{widetext}
$\tilde{R}_{\alpha}$ satisfies the 1D equation
\begin{equation}\label{ksStar1dreduced}
  \tilde{\mathcal H}^*_{m_\alpha}(\bar{r})\tilde{R}_{\alpha}(\bar{r})=\epsilon_{\alpha}\tilde{R}_{\alpha}(\bar{r}),
\end{equation}
where $\bar{r}=r/l_B$ and
\begin{eqnarray}\label{1DksStarpolarform}
  &&\tilde{\mathcal H}^*_{m_\alpha}(\bar{r})=V^*_{\rm KS}+\nonumber\\
  \frac{1}{2}&& \hbar \omega^*_B\bigg[-\frac{\partial^2}{\partial\bar{r}^2}+\frac{m_\alpha^2-\frac{1}{4}}{\bar{r}^2}-m_\alpha\frac{\mathcal{B}(\bar{r})}{B}
  +\frac{\mathcal{B}^2(\bar{r})}{B^2}\frac{\bar{r}^2}{4}\bigg]\;.
\end{eqnarray}
Eq.~(\ref{ksStar1dreduced}) can be solved numerically for each angular momentum $m_\alpha\neq 0$ using the finite-difference method. 

For $m=0$, the finite difference method does not perform well because a singular point arises at $\vec{r}=0$ due to the second term in the rectangular bracket in Eq.~(\ref{1DksStarpolarform}).  For $m=0$, we alternatively solve Eq.~(\ref{singleCFKS}) using a basis expansion as follows. The matrix form of $\mathcal H^*$ can be obtained under the basis set $\textbf{H}_0=\{\phi_{n,m=0}({\bar{\vec{r}}};l_{\mathcal{B}_0}), n=0,1,\ldots,N_{\rm L}\}$ in the angular momentum $m=0$ subspace, where the basis $\phi_{n,m}({\bar{\vec{r}}};l_{\mathcal{B}_0})$ is defined in Eq.~(\ref{LandauWave}) and $N_{\rm L}$ is the $\Lambda$L cutoff. We discuss below how we fix  $l_{\mathcal{B}_0}$.

Using a dimensionless form $\bar{\vec{r}}=\vec{r}/l_{\mathcal{B}_{m}}$, we are now able to write down the matrix element $\mathcal{H}^*_{(n',n)}$ of $\mathcal{H}^*$ in the subspace $\textbf{H}_0$,
\begin{eqnarray}\label{Kineticdim}
&&\mathcal{H}^*_{(n',n)}=T^{*}_{(n',n)}+V^{*}_{\rm{KS}{(n',n)}},\\
&& T^{*}_{(n',n)}=\langle\phi_{n',0}(\bar{\vec{r}};l_{\mathcal{B}_0})|T_0+V_0|\phi_{n,0}(\bar{\vec{r}};l_{\mathcal{B}_0})\rangle,\\
 &&=\hbar \omega_{\mathcal{B}_0}(n+\frac{1}{2})\delta_{n'n}+\langle\phi_{n',0}(\bar{\vec{r}};l_{\mathcal{B}_0})|V_0|\phi_{n,0}(\bar{\vec{r}};l_{\mathcal{B}_0})\rangle,\\
&&V^{*}_{{\textrm{KS}}{(n',n)}}=\langle\phi_{n',0}(\bar{\vec{r}};l_{\mathcal{B}_0})|V^{*}_{\rm{KS}}(\bar{\vec{r}})|\phi_{n,0}(\bar{\vec{r}};l_{\mathcal{B}_0})\rangle,
\end{eqnarray}
where
\begin{eqnarray}
& &T_0 = \frac{1}{2}\hbar \omega_{\mathcal{B}_0}[\nabla_{\bar{\vec{r}}}^2+\frac{\hat{L}_{\bar{z}}}{\hbar}+\frac{\bar{r}^2}{4}], \label{KineticPartsdim1}\\
& &V_0 = \frac{1}{2}\hbar \omega_{\mathcal{B}_0}[\frac{\mathcal{B}(\bar{r})-\mathcal{B}_{0}}{\mathcal{B}_{0}}\frac{\hat{L}_{\bar{z}}}{\hbar}+ \frac{1}{4}\frac{\mathcal{B}^2(\bar{r})-\mathcal{B}^2_{0}}{\mathcal{B}^2_{0}}\bar{r}^2],\label{KineticPartsdim2}
\end{eqnarray}
with $\omega_{\mathcal{B}_0}=\frac{e\mathcal{B}_{0}}{m^*c}$. After diagonalizing the $(N_{\rm L}+1)\times(N_{\rm L}+1)$ square matrix of $\mathcal{H}^*_{(n',n)}$, we will obtain the single CF orbital spanned as a linear superposition as:
\be
\psi_{n,0}(\vec{r})=\sum_{n'=0}^{N_{\rm L}} c_{n',0}\phi_{n',0}(\vec{r}, l_{\mathcal{B}_0}),
\ee
as well as its eigenenergy $\epsilon_{\{n,0\}}$.

A reasonable choice for $\mathcal{B}_0$ (which determines $l_{\mathcal{B}_{0}}$) assures rapid convergence. We find it convenient to choose $\mathcal{B}_0$ by solving
\begin{equation}\label{m0approxiamtion}
\hbar\frac{e\mathcal{B}_{0}}{m^*c} =\frac{\Delta_{m=-1}+\Delta_{m=1}}{2},
\end{equation}
where $\Delta_{m=\pm1}$ is the mean value of cyclotron energy gap between the lowest two energy levels for the $m=\pm 1$ angular momentum subspace, obtained using the finite difference method. This is justified because the orbitals with $m=0$ are spatially overlapping with the $m=\pm1$ orbitals, and thus experience a similar effective magnetic field. We fix $N_{\rm L}=19$ for calculations in this paper.

In the presence of the KS potential and a non-uniform effective magnetic field, the eigenstate of $\mathcal{H}^*$ mixes among the unperturbed $\Lambda$Ls of composite fermions. The solutions to the Eq.~(\ref{singleCFKS}) form the renormalized $\Lambda \rm L$s. Each eigenstate $\psi_{\alpha}$ can be labeled by quantum numbers $\{n,m\}$. With the above method, we can obtain the self-consistent solution of Eq.~(\ref{singleCFKS}) using the following iterative procedure. (i) We start with a set of input KS orbitals, which generates the input density $\rho_{\rm in}$. (ii) We obtain $T^*$ and $V^*_{\rm KS}(\vec{r})$ on the left-hand side of Eq.~(\ref{singleCFKS}), diagonalize the Hamiltonian to obtain the KS orbitals, and occupy the lowest $N$ KS orbitals to obtain the output density $\rho_{\rm out}=\sum_{n,m} c_{n,m}|\psi_{n,m}|^2$, where $c_{n,m}$ is the occupation number for KS orbitals labeled by $\{n,m\}$. (iii) We determine the relative difference $\frac{\Delta N}{N}$ between the input and output $\rho$, where $\Delta N=\int |\rho_{\rm in}-\rho_{\rm out}|d^2 \vec{r}$. We accept $\rho_{\rm out}$ as converged if $\frac{\Delta N}{N}<0.1\%$ is satisfied. (iv) If $\rho$ is not converged, we prepare new input density $\rho_{\rm in}$ by mixing some output density into the previous input: $\rho_{\rm in}\rightarrow \eta\rho_{\rm in}+(1-\eta)\rho_{\rm out}$, where the mixing coefficient is $\eta\geq 0.95$ in this paper. We iterate the process, until convergence is reached.

\section{Importance of self-consistency}\label{densityComparisonSec}

\begin{figure}[t]
\includegraphics[width=\columnwidth]{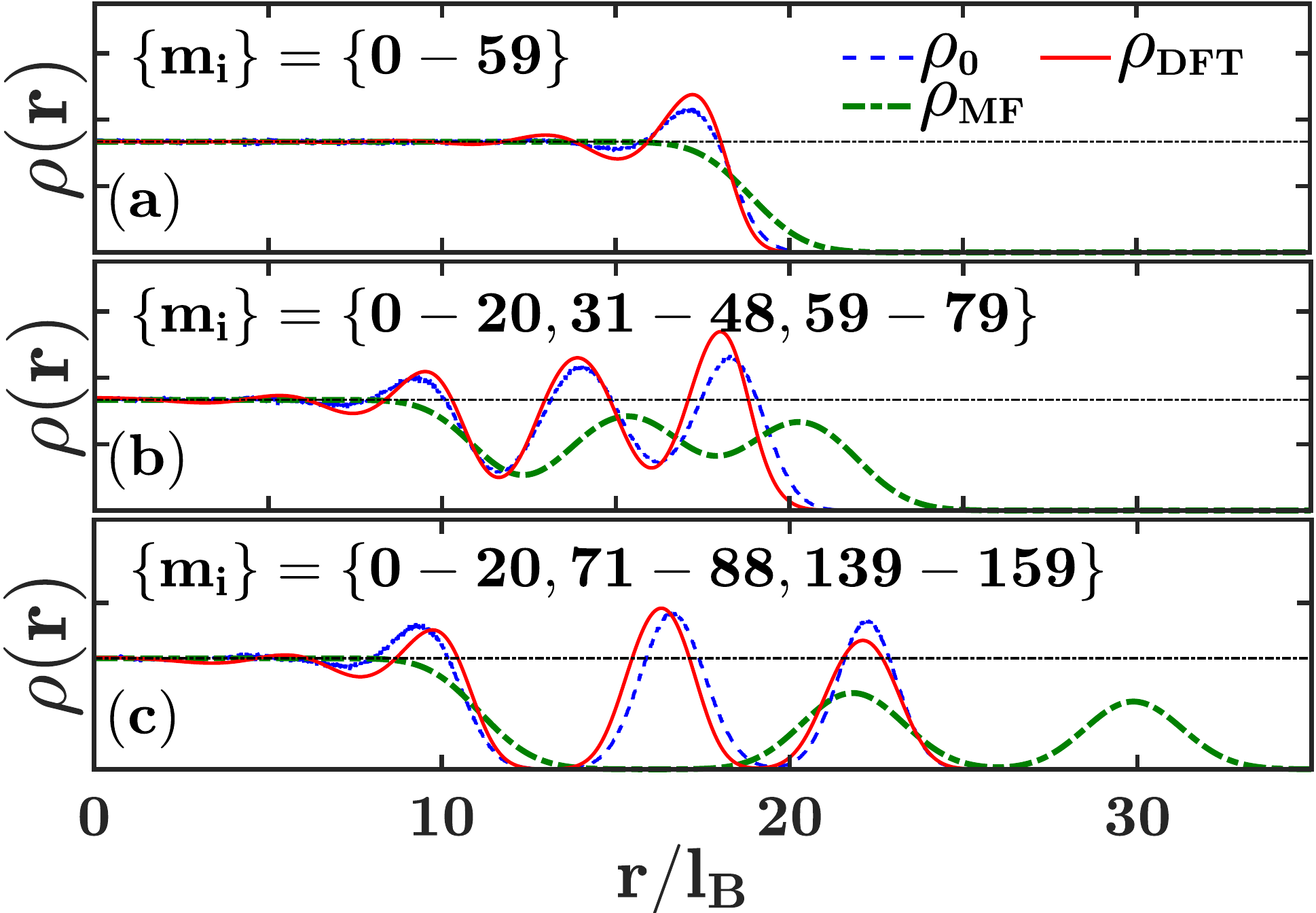}
\caption{We consider three configurations for composite fermions: $\{  m_i \}=\{0-59\}$, $\{  m_i \}=\{0-20, 31-48, 59-79\}$ and $\{  m_i \}=\{0-20, 71-88, 139-159\}$.  The electron density is $\rho_{0}(\vec{r})$ is obtained from the explicit electron wave function. The density $\rho_{\rm MF}$ is obtained from a mean-field approximations, explained in the text, which assumes composite fermions at a fixed effective magnetic field. The density profile obtained from CFDFT is labeled as $\rho_{\rm DFT}$. All densities are quoted in units of the density at $\nu=1$.}\label{MCDFTcompare}
\end{figure}

Let us consider a configuration of composite fermions occupying angular momenta $\{  m_i \}$. The explicit electron wave function for this state is given by 
\be
\Psi_0={\rm Det}[ z_i^{m_j} ] \prod_{k<l}(z_k-z_l)^2 \exp[-\sum_j |z_j|^2/4l_B^2]
\ee
The density obtained from this wave function, labeled $\rho_0$, is shown in Fig.~\ref{MCDFTcompare} for three different configurations: $\{  m_i \}=\{0-59\}$, $\{  m_i \}=\{0-20, 31-48, 59-79\}$ and $\{  m_i \}=\{0-20, 71-88, 139-159\}$. We note that these states are not ground states but are considered for illustration.

In a mean field (MF) approximation, we consider a model of composite fermions at a constant effective magnetic field with $B^*=B/3$. Its wave function is given by
\be
\Psi_{\rm MF}=\det[z_j^{m_i}\exp(-|z_j|^2/4l^{*2})]
\ee
where $l^{*}=\sqrt{3}l_B$. The resulting density $\rho_{\rm MF}(\vec{r})$ is also given in Fig.~\ref{MCDFTcompare} and seen to be a poor approximation of $\rho_0$. In particular, it fails to obtain the correct positions of the peaks or  the overshoot of the densities above the bulk value near the edge. Incidentally, it is also evident that the MF approximation does not capture the physics of fractional statistics, because the position of a quasihole in the angular momentum $m$ orbital does not depend on whether it encloses any other quasiholes.

This figure also shows the density calculated from constrained CFDFT, in which we fix the angular momenta of composite fermions but obtain the wave functions of the orbitals by a self consistent solution of the KS equations. The self-consistent densities, labeled $\rho_{\rm DFT}$, are in excellent agreement with $\rho_0$. These comparisons further underscore the non-triviality of CFDFT and the importance of the self-consistency in the KS equation that ensures that the magnetic field $B^*(\vec{r})$ is consistent with the local density.

\section{Fractional braid statistics and background density fluctuations}\label{RecalibratedStatistics}

\begin{figure}[t]
\includegraphics[width=\columnwidth]{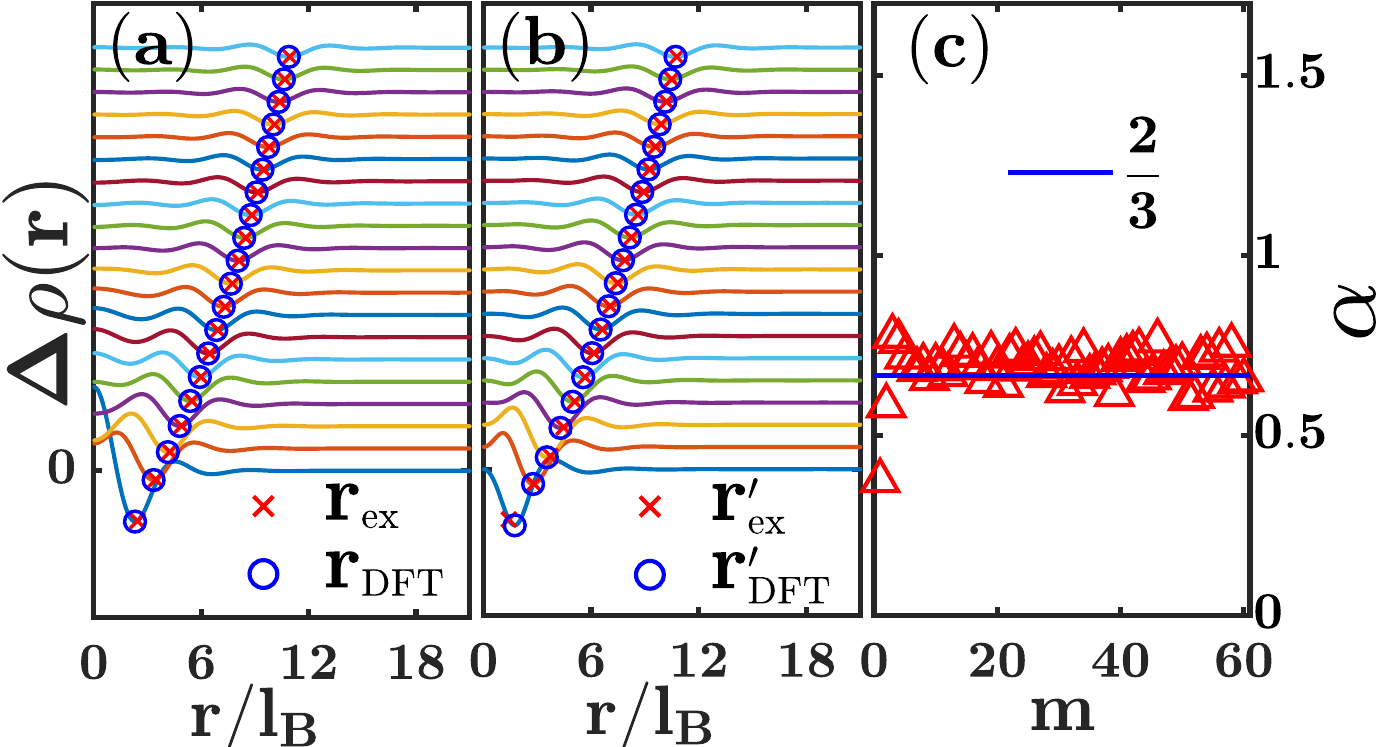}
\caption{{\bf Fractional braid statistics revisited.} Panel {\bf (a)} shows the electron density difference for a system before and after a quasihole in angular momentum $m$ orbital is created, with $m$ changing from 1 to 20 for the curves from the bottom to the top. (Each successive curve has been shifted up vertically for clarity.) Panel {\bf(b)} shows the same in the presence of another quasihole at the origin. For each $m$, we indicate the expected position of the outer quasihole (red cross) as well as the position obtained from the DFT density determined by locating the local minimum of in the density difference (blue circle). Panel {\bf(c)} shows the statistics parameter $\alpha\equiv (r^2_{\rm{\rm DFT}}-r^{\prime 2}_{\rm{\rm DFT}})/6l_B^2 $. The calculation has been performed for $N=200$ composite fermions at $\nu_0=1/3$. All parameters are the same as that in Fig.~3 in the main paper.}\label{BraidingSI}
\end{figure}

In Fig.3 of the main paper, we commented that deviations in the statistical parameter $\alpha$ from the expected value arise because the density of the background FQHE liquid droplet also shows oscillations (due to finite system size), which causes a slight shift in the position of the local minimum where the outer quasihole resides. To eliminate the effect of the density oscillations to the best extent possible, we calculate the density with and without the outer quasihole, and use the {\it difference} between the two densities to determine its position $\vec{r}_{\rm DFT}$. An analogous calculation in the presence of the additional quasihole at the origin yields $\vec{r}'_{\rm DFT}$. The resulting statistics parameter $\alpha$ is much better quantized at $2/3$, as seen in Fig.~\ref{BraidingSI}. Interestingly, the quantized value persists to very small quasihole separations.


\end{document}